\documentstyle[12pt,epsfig]{article}
\def\simlt{\stackrel{<}{{}_\sim}}

\def\be{\begin{equation}}
\def\ee{\end{equation}}
\def\bea{\begin{eqnarray}}
\def\eea{\end{eqnarray}}
\def\simlt{\stackrel{<}{{}_\sim}}

\def\ap#1#2#3   {{\em Ann. Phys. (NY)} {\bf#1} (#2) #3.}
\def\apj#1#2#3  {{\em  Astrophys. J.} {\bf#1} (#2) #3.}
\def\apjl#1#2#3 {{\em Astrophys. J. Lett.} {\bf#1} (#2) #3.}
\def\app#1#2#3  {{\em Acta. Phys. Pol.} {\bf#1} (#2) #3.}
\def\ar#1#2#3   {{\em Ann. Rev. Nucl. Part. Sci.} {\bf#1} (#2) #3.}
\def\cpc#1#2#3  {{\em Computer Phys. Comm.} {\bf#1} (#2) #3.}
\def\err#1#2#3  {{\em Erratum} {\bf#1} (#2) #3.}
\def\ib#1#2#3   {{\em ibid.} {\bf#1} (#2) #3.}
\def\jmp#1#2#3  {{\em J. Math. Phys.} {\bf#1} (#2) #3.}
\def\ijmp#1#2#3 {{\em Int. J. Mod. Phys.} {\bf#1} (#2) #3.}
\def\jetp#1#2#3 {{\em JETP Lett.} {\bf#1} (#2) #3.}
\def\jpg#1#2#3  {{\em J. Phys. G.} {\bf#1} (#2) #3.}
\def\mpl#1#2#3  {{\em Mod. Phys. Lett.} {\bf#1} (#2) #3.}
\def\nat#1#2#3  {{\em Nature (London)} {\bf#1} (#2) #3.}
\def\nc#1#2#3   {{\em Nuovo Cim.} {\bf#1} (#2) #3.}
\def\nim#1#2#3  {{\em Nucl. Instr. Meth.} {\bf#1} (#2) #3.}
\def\np#1#2#3   {{\em Nucl. Phys.} {\bf#1} (#2) #3.}
\def\pcps#1#2#3 {{\em Proc. Cam. Phil. Soc.} {\bf#1} (#2) #3.}
\def\pl#1#2#3   {{\em Phys. Lett.} {\bf#1} (#2) #3.}
\def\prep#1#2#3 {{\em Phys. Rep.} {\bf#1} (#2) #3.}
\def\prev#1#2#3 {{\em Phys. Rev.} {\bf#1} (#2) #3.}
\def\prl#1#2#3  {{\em Phys. Rev. Lett.} {\bf#1} (#2) #3.}
\def\prs#1#2#3  {{\em Proc. Roy. Soc.} {\bf#1} (#2) #3.}
\def\ptp#1#2#3  {{\em Prog. Th. Phys.} {\bf#1} (#2) #3.}
\def\ps#1#2#3   {{\em Physica Scripta} {\bf#1} (#2) #3.}
\def\rmp#1#2#3  {{\em Rev. Mod. Phys.} {\bf#1} (#2) #3.}
\def\rpp#1#2#3  {{\em Rep. Prog. Phys.} {\bf#1} (#2) #3.}
\def\sjnp#1#2#3 {{\em Sov. J. Nucl. Phys.} {\bf#1} (#2) #3.}
\def\spj#1#2#3  {{\em Sov. Phys. JEPT} {\bf#1} (#2) #3.}
\def\spu#1#2#3  {{\em Sov. Phys.-Usp.} {\bf#1} (#2) #3.}
\def\zp#1#2#3   {{\em Zeit. Phys.} {\bf#1} (#2) #3.}

\def\beq{\begin{equation}}
\def\eeq{\end{equation}}
\def\bea{\begin{eqnarray}}
\def\eea{\end{eqnarray}}

\def\k{m_{eff}}
\begin{document}
\begin{flushright}
\vspace{-0.3cm}
{\normalsize CERN-TH/99-173}\\ 
{\normalsize ANL-HEP-PR-99-76}\\
\end{flushright}
\vspace{1.cm}
\begin{center}
{\Large \bf Neutrino Masses, Mixing Angles and the\\
\vskip 0.5cm
Unification of Couplings in  the MSSM  
%
%
}\\ 
~\\
~\\
~\\
{\bf M. Carena$^{a,\dagger}$,
J. Ellis$^a$, S. Lola$^a$ and C.E.M. Wagner$^{a,b}$} \\
 ~\\
~\\
~$^{a)}$ Theory Division, CERN, CH 1211, Geneva, Switzerland. \\
~\\
~$^{b)}$ Argonne National Laboratory, 9700 Cass Ave., Argonne, IL 60439,
USA
~\\
~\\
~\\
\end{center}
\begin{quote}
\begin{center}
{\bf Abstract}
\end{center}
In the light of the gathering evidence for $\nu_{\mu}-\nu_{\tau}$
neutrino oscillations, coming in particular from
the Super-Kamiokande data on  atmospheric neutrinos,
we re-analyze the unification 
of gauge and Yukawa couplings within the minimal supersymmetric
extension of the Standard Model (MSSM).
Guided by a range of different grand-unified models, we stress
the relevance of large mixing in the lepton sector for the question
of bottom-tau Yukawa coupling unification.
We also discuss the dependence of the favoured value of $\tan\beta$ on
the characteristics of the high-energy quark and lepton mass matrices.
In particular, we find that, in the presence
of large lepton mixing, Yukawa unification
can be achieved for intermediate values of $\tan\beta$ that were
previously disfavoured. The renormalization-group sensitivity to the
structures of different mass matrices may enable Yukawa
unification to serve as a useful probe of GUT models.
\end{quote}
\vskip 1.0cm 
\begin{flushleft} 
{\normalsize CERN-TH/99-173}\\
\end{flushleft} 
~\\
$^{\dagger}$ {\small On leave of
absence from Fermi National Accelerator Laboratory, Batavia, IL 60510, USA} 

\setcounter{footnote}{0}
\newpage 
\noindent


\section{Introduction}

The most mysterious aspect of the Standard Model may be the pattern of fermion
masses and mixing. These are characterized by many parameters, whose patterns
are difficult to discern. Two major classes of theoretical approaches 
to the question of fermion masses
may be distinguished: Grand Unified Theories (GUTs), whose non-Abelian gauge
symmetries may relate the masses of fermions in the same generation via
Clebsch-Gordan coefficients of order unity, that are renormalized in a
calculable way, and global flavour symmetries such as U(1), which may explain
the observed hierarchies of fermion masses and mixing between different
generations. The first (and only successful) GUT mass prediction was that $m_b
= m_\tau$ before renormalization~\cite{CEG}, which is compatible with the
measured values
at the $\sim$ 10--30\% level, in both the Standard Model and the 
Minimal Supersymmetric
Standard Model (MSSM)~\cite{susymb}. Optimal
agreement within the MSSM is found either for $\tan\beta \le 2$ or for
$\tan\beta\ge 30$, with the attractive possibility of $b - \tau - t$ Yukawa
coupling
unification in the latter case~\cite{btau,CPW,btaut}. The na\"\i ve parallel
predictions $m_s =
m_\mu$ and $m_d = m_e$ are unsuccessful, but these may be modified by
non-trivial Clebsch-Gordan coefficients~\cite{factor3} and/or
higher-dimensional contributions~\cite{slop}
to the fermion mass matrices controlled by global U(1) flavour symmetries.

Important new information is now being provided by the emerging pattern of
neutrino masses and mixing~\cite{SKam,KamMac}. Various predictions for
these had been made in the
context of different models for quark and lepton
masses~\cite{textures,neutr}, which are now being
winnowed out by the experimental measurements. In particular, 
the possibility that the Maki-Nakagawa-Sakata (MNS)~\cite{mns} lepton
mixing angles might be large
had not always been anticipated, and raise questions about the
viability of the successful prediction $m_b = m_\tau$.
The purpose of this paper is to explore the circumstances under which this
prediction can be retained, based on a study of generic patterns of mixing
among neutrino and charged-lepton flavours, and to understand the corresponding
flexibility in the range of $\tan\beta$ compatible with $b-\tau$ Yukawa
unification and other hypotheses. 

Our work has been stimulated by reports from the
Super-Kamiokande Collaboration~\cite{SKam}, which
are also supported by other experiments~\cite{KamMac}, 
that confirm previous measurements that the 
$\nu _{\mu }/\nu _{e }$ ratio
in atmospheric neutrinos is 
smaller than the Standard Model expectations. 
The existing data favour~\cite{SKam,chooz}
$\nu _{\mu }$--$\nu _{\tau }$ oscillations, 
with the following constraints on the
mass differences  and the mixing: 
\begin{eqnarray}
\delta m_{\nu _{\mu }\nu _{\tau }}^{2} &\approx &(10^{-2}\;{\rm to}%
\;10^{-3})\;{\rm eV^{2}} \\
\sin^{2}2\theta _{\mu \tau } &\geq &0.9  \label{atmos}
\end{eqnarray}
%

If one departs from the Standard Model and
assumes the existence of non-zero neutrino
masses, it is natural also to resolve
the solar-neutrino deficit through neutrino
oscillations. In the case of vacuum oscillations,
one requires large mixing and a splitting between the oscillating
neutrinos in the range
$\delta m_{\nu _{e}\nu _{\alpha }}^{2} 
\approx (0.5-1.1)\times 10^{-10}~{\rm eV}^2$,
where $\alpha $ is $\mu $ or $\tau$. On the other hand,
matter-enhanced oscillations~\cite{MSW} within the Sun would allow for
both
small
and large mixing, with
$ \delta m_{\nu _{e}\nu _{\alpha }}^{2} \approx (0.3-20) \times 10^{-5}
~{\rm eV}^2$~\footnote{We do not pursue here evidence from the  LSND
Collaboration 
for ${\bar \nu}_{\mu }$--$\bar{\nu}_{e}$ and ${\nu }_{\mu }$--${\nu }_{e}$
oscillations \cite{LSND2}, which has not yet been
confirmed by the KARMEN 2 experiment~\cite{Karmen2}.}.
%

The most straightforward extension of the Standard Model
that one may consider is to 
include three new right-handed neutrino states, with Yukawa
couplings similar to those of the other fermions.
The easiest way to explain the smallness of the neutrino masses is 
then to 
assume that the right-handed neutrinos acquire a gauge-invariant
but lepton-number-violating Majorana mass of the type
$M_{\nu_R} \nu _{R} \nu_R$, with
$M_{\nu_R} \gg M_W$.
Neutrino masses are then determined
%
%
via a see-saw mechanism~\cite{seesaw}, with the light neutrino 
eigenvalues being determined by
the diagonalization of the matrix 
\begin{equation}
M = \left( \begin{array}{cc}
0 & m^D_\nu\\
m_\nu^{D^{T}} & M_{\nu_R}
\end{array}
\right)
\label{seesaw}
\end{equation}
to be
$ m_{light} \simeq \frac{(m_\nu^{D})^{2}}{M_{\nu_R}} $,
where $m_\nu^{D}$ is the Dirac neutrino mass matrix,
and hence naturally suppressed.

Since both the solar and atmospheric
deficits  require small mass differences, 
there are two 
possible neutrino hierarchies that could explain them
simultaneously:

a) Textures with almost 
degenerate neutrino eigenstates, with mass ${\cal O} 
({\rm eV})$.  In this case neutrinos may
also provide astrophysical hot dark matter, and

b) Textures with 
large hierarchies of neutrino masses:
$m_{3} \gg m_{2} \gg m_{1}$, in which case
the atmospheric neutrino data require
$m_{3} \approx (0.03 \; {\rm to} \;
0.1)$ eV
and $m_{2} \approx (10^{-2} \; {\rm to} \;
10^{-3})$ eV. 

In this article, we shall concentrate on this second
possibility, motivated in part by the difficulties thrown up
by renormalization effects in models with degenerate
neutrinos~\cite{ELLN3,CEIN}.

As we have emphasized, one of the major challenges in
high-energy physics is the origin of the fermion
masses and mixing angles, and the neutrino data are now providing us
with precious new information. It is interesting to investigate
if the pattern of neutrino masses and mixing angles may be accomodated
in a natural way within a GUT scenario~\cite{gut,HO,gui1,MG}, 
with supersymmetry realized at
low energies. Indeed, the successful unification of gauge couplings
provides one of the main experimental motivations for low-energy
supersymmetry. For values of $\alpha_3(M_Z) \simeq 0.118$,
gauge-coupling unification is accurate for a supersymmetric spectrum
with masses of
the order of 1 TeV, whereas, as we said above,
in the absence of neutrino masses,
$b-\tau$ mass unification demands either small values of $\tan\beta$,
close to the infrared fixed-point solution of the MSSM
renormalization-group equations, or
large values of $\tan\beta$~\cite{LP,CPW,Nir2,BCPW}.

In the presence of neutrino masses,
the running of the various couplings
from the unification scale
down to low energies is modified. From $M_{GUT}$ to the scale $M_N$, at
which
the effective light-neutrino mass operator appears, 
the effects of the neutrino Yukawa coupling essentially
cancel those of the top Yukawa coupling in the $m_b/m_{\tau}$
ratio, if one assumes
unification of the third-generation neutrino and top Yukawa couplings.
This makes
precise unification hard to achieve~\cite{VS,Rat}. 
However, it was shown in \cite{LLR} that precise
unification can be restored in the presence of large lepton
mixing.  Since the right handed neutrinos are neutral,
the unfication of gauge couplings is only
affected by the presence of the Dirac 
Yukawa coupling at the two-loop level. The largest 
effects are obtained for values
of the neutrino masses such that the 
neutrino Yukawa coupling becomes strong, close to the
limit of perturbativity, at scales of the order of $M_{GUT}$.
The effect of the neutrino Yukawa coupling lowers
the predicted value of strong gauge
coupling, $\alpha_3(M_Z)$, by less than
one percent. Below, we make an extensive study of the 
unification of couplings, taking into account
the information available from
Super-Kamiokande and other neutrino experiments.

Below $M_N$, the $\nu_R$ decouple from the 
spectrum, and the quantity that gets 
renormalized is an effective  
neutrino operator of the type $\nu_L \nu_L H H$~\cite{Bab,run}.
Renormalization-group
effects, which may be summarized by simple semi-analytic
expressions both in the ranges
$M_{GUT}$ to $M_N$~\cite{LLR} and $M_N$ to low 
energies \cite{ELLN2,Hab,ELLN3}, may give important information on
the structure of the neutrino textures. 
Different boundary  conditions for the Yukawa 
couplings of the third generation of quarks and leptons at $M_{GUT}$ and 
the precise 
value of $M_N$ may constrain, for different ranges of $\tan \beta$ 
\cite{LLR,ELLN3,CEIN},  the amount of 
mixing in the lepton sector. Quite generically, a given 
amount of mixing at the GUT scale 
may be amplified  or destroyed  at low energies, and vice versa, due to strong 
renormalization-group effects.

%
%
%

In the present work, we analyze in a systematic
way the implications of the recent experimental information on
neutrino masses and mixing for the unification of couplings at a high
energy scale. We shall concentrate on the small 
and moderate $\tan\beta$ regime, 
in which the impact of the lepton mixing on the question of unification
of coupling becomes most relevant.
We review in Section 2 our theoretical framework for analyzing fermion
textures,  which is a generic supersymmetric $SO(10)$ GUT with
higher-dimensional
interactions. We use this to establish examples of different possible origins
of large neutrino mixing. In Section 3, we analyze the renormalization-group
running of quark, lepton and neutrino masses below the GUT scale $M_{GUT}$,
both above and below the intermediate seesaw mass scale $M_N$, establishing the
conditions for maximal neutrino mixing. In Section 4, we present numerical
results which show how the successful prediction $m_b = m_\tau$ may be
retained, even for intermediate values of $\tan\beta$ that were previously
disfavoured in the MSSM. 
In section 5 we analyse the results of
using the precise $SU(5)$ relations between the mixing of the left-handed
leptons and the right-handed down-quark states. 
Finally, in Section 6 we summarize our conclusions and
identify outstanding issues for future study.

%
%

\section{Theoretical Framework for Fermion Textures}

We now discuss examples how the classes of
mass matrices that we study, embodying
bottom-tau unification at the GUT scale,
may be realized. As already stressed, 
the hierarchical structure  of the fermion mass matrices 
suggests that they might be generated by 
an underlying family symmetry.
The various Standard Model fields would have
certain charges under this symmetry, and a coupling
represented by a given operator is allowed
if its net flavour charge is zero. 
Since the top, bottom and $\tau$ masses are larger than
the rest of the fermion masses with the same quantum numbers, a natural
expectation is that they arise via renormalizable terms.
The remaining entries are hypothesized to be generated 
from higher-dimensional operators when the
flavour symmetry is spontaneously broken, via vev's for
fields that are singlets of the
Standard Model gauge group,
with non-trivial 
flavour charge. 
Precise unification of the bottom and tau masses may 
take place if we work within a GUT scheme, which may lead to
specific Clebsch-Gordan
relations between the couplings of quarks and leptons.
Since, it is important for our purposes to know
the exact relation between $\lambda_b(M_{GUT})$ and 
$\lambda_\tau(M_{GUT})$,
we now describe a framework where
exact $b-\tau$ unification is guaranteed, whilst the mixing in the
left-handed lepton sector may be quite different from that
in the right-handed down-quark sector.

As a specific example of a GUT symmetry that
unifies quarks and leptons in common representations,
leading to various relations
between the charges of the quark and lepton fields,
and hence to relations between their mass
hierarchies, we consider $SO(10)$ models~\footnote{Analogous
studies could be made in the context of flipped $SU(5) \times
U(1)$: see~\cite{ELLN2} and references therein.}.
A feature of this model approach is 
that one can also account for the mass
splitting within a particular family.
Realistic mass matrices
may be obtained by
considering the effects of the Higgs multiplets
that are necessary  in order to break $SO(10)$ to
$SU(3) \times SU(2) \times U(1)$ \cite{ADH10}.
The smallest such
representations are ${\bf 45}$, ${\bf 16}$ 
and ${\bf \overline{16}}$, which may be used to
generate operators with rank $\geq 4$ in the
mass matrices. 
Along these lines, one can use
the $SO(10)$-invariant Yukawa interaction
$O_{33} =  A  \cdot {\bf 16}_3\ {\bf 10}\ {\bf 16}_3$ to
give mass to the fermions of the third family
in terms of a single coupling $A$.
Higher-order operators are of the general type
\begin{equation}
O_{ij} = {\bf 16}_i  \;\;  {M_{GUT}^k ~{\bf 45}_{k+1} \cdots {\bf 45}_m
\over M_P^l
{}~{\bf 45}_X^{m-l}} \;\;
  {\bf 10}  \;\; {M_G^n ~{\bf 45}_{n+1} \cdots {\bf 45}_p \over M_P^q
{}~{\bf 45}_X^{p-q}} \;\;
 {\bf 16}_j 
\end{equation}
where the {\bf 45} representations in the numerator are along any of the
four
directions $X,Y,B-L,T_{3R}$ \footnote{
In particular, the ${\bf 45}$
is the adjoint representation of $SO(10)$, and
therefore its vev may point in any direction
in the space of the 45 generators of $SO(10)$,
as long as it leaves the standard model gauge
group unbroken.  
A vev in the $X$  direction
is required for the breakdown of $SO(10)$ to $SU(5) \times U(1)_X$
at a scale $M_{10}$ between $M_{GUT}$ and $M_P$.
The other directions are associated with the
breaking of $SU(5)$ down to the Standard Model gauge group.}.
The mass terms of the
first and second fermion families are generated 
from particular non-renormalizable
operators, whose coefficients are suppressed by a set of
large scales. Whether the mass matrices will be symmetric
or non-symmetric depends on the choice of operators.
We concentrate in this paper on models with symmetric mass matrices,
since these are sufficient to illustrate our key points: our
analysis could easily be extended to include non-symmetric cases.

Each direction in $SO(10)$ is associated with
different quantum numbers for different family members,
which can give different but 
related coefficients between the quark
and lepton mass matrices.
For completeness, we present  these coefficients 
in Table 1.
As we see from the Table 1,
the absolute values of 
the coefficients of interest vary from 0 to 6, indicating 
that we can in principle obtain very different hierarchies
between the textures of quark and lepton masses. 
Taking into account the known low-energy data,
acceptable models have been identified 
\cite{ADH10,CMS}, within which
the lower $2 \times 2$ up-, down-quark and charged-lepton 
mass matrices  are written as
\begin{eqnarray}
m_{U,D,E} \propto  \left( 
\begin{array}{c c }
 y_a E e^{i\phi} & x^{'}_a B \\
 x_a B & A 
\end{array}
\right),
\end{eqnarray}
where $x_a$, $x^{'}_a$ and $y_a$ are Clebsch-Gordan factors, whilst
$A$, $B$, $E$ and $\phi$  are 
arbitrary parameters, with $A \gg B,E$ being
adjusted by the fermion data.
These textures can
be generated both in the large~\cite{ADH10}
and small~\cite{CMS}
$\tan\beta$ regimes (for the latter case, see
Appendix A).

\begin{table}[tbp]
\begin{center}
\begin{tabular}{|c|c|c|c|c|}
\hline
&$X$ &$ Y$&$B-L$&$T_{3R}$\cr
\hline
  $q$ &1 &1 & 1 &0\cr
$u^c$  &1 &-4&-1&1\cr
  $d^c$&-3 &2&-1&-1\cr
$\ell$&-3&-3&-3  &0\cr
$e^c$  &1&6 &3  &-1\cr
  $\nu^c$&5 &0&3  & 1\cr
\hline
\end{tabular}
\end{center}
\caption{\it $X,Y,B-L$ and $T_{3R}$ quantum numbers of the 
Standard Model fermions}
\end{table}

Following the Super-Kamiokande data,
we require in addition 
large mixing in the (23) lepton sector,
while the (23) quark mixing is small. 
Moreover, we need 
$\frac{m_{\mu}}{m_{\tau}} >
\frac{m_{s}}{m_{b}}$,
as well as certain cancellations between
the various entries in the lepton sector,
in order to obtain the correct 
$\frac{m_{\mu}}{m_{\tau}}$ ratio. There is some  freedom
to achieve the desired cancellations
by an appropriate choice of $E$ and
$y_a$, however the chosen operators have
also to match the quark data.
In what follows, we will try to identify
some viable examples,
without doing a complete operator analysis.

There are in principle two ways to obtain the
correct lepton and quark contributions simultaneously:

(i) The easiest way to search for suitable operators
is to assume that the operators contributing to the
(23) and (32) elements are such that the Clebsch-Gordan
factors controlling the lepton mass matrix elements
are much larger than the quark ones.  
These operators are then assumed to lead to 
the relevant lepton entries, whilst the associated up- and 
down-quark  terms could come either from these same operators or 
from other subleading ones. 

(ii) One can add two different operators, in a way
that cancellations are achieved. The difficulty
in this case is that the vev's and coefficients of two different 
operators need to match quite well in order to lead to the desired
results.

We first concentrate on the measured (23) lepton
mixing. This can arise either from the charged-lepton sector,
or the neutrino sector, or both. 
Indeed, the leptonic mixing matrix
$V_{MNS}$ of Maki, Nakagawa and Sakata~\cite{mns} is defined
in a way similar to the Cabibbo-Kobayashi-Maskawa mixing
matrix $V_{CKM}$ for the quark currents:
\bea
\label{vmns}
V_{MNS} =
V_{\ell}V_{\nu}^{\dagger}
\eea
where $V_{\ell}$ transforms the left-handed charged leptons
to a diagonal mass basis, whereas
$V_{\nu}$ diagonalizes the light-neutrino mass matrix~\cite{seesaw},
\begin{equation}
m_{eff}
=m^D_{\nu}\cdot (M_{\nu_R})^{-1}\cdot m^{D^{\normalsize T}}_{\nu}.
\label{eq:meff}
\end{equation}
In the above
$m^D_{\nu}$ is the Dirac neutrino mass matrix and
$M_{\nu_R}$ the heavy Majorana mass matrix.
We assume that
the lepton mass hierarchy originates from the structure
of the Dirac neutrino and charged-lepton mass matrices.
Quite generally, under these conditions, 
the structure of the heavy Majorana mass
matrix $M_{\nu_R}$ does not affect the low-energy lepton mixing (see
Appendix B).
We further concentrate on the case (i) discussed above, looking for
operators of relatively low order, so that they are
not suppressed by the high scales of the theory.
For simplicity of presentation, we impose the requirement of 
symmetric mass matrices, which
constrains the possible choices.

We remark that the lower-order symmetric operators 
with Clebsch-Gordan factors with large differences 
between quarks and leptons tend to lead to similar mixing
angles in the charged-lepton and neutrino sectors, and
hence to small lepton mixing. This is, e.g., the case with the
operator
\begin{equation}
{\bf 16} \; {\bf 4}5_{B-L} \; {\bf 10} \; {\bf 45}_{B-L} \; {\bf 16},
\end{equation}
which leads to Clebsch-Gordan factors $x_u = x_d = -1$ and $x_{\ell} = 
x_{\nu} = -9$. On the other hand, there also exist
operators that lead to large differences between the individual
quark and lepton Clebsch-Gordan coefficients, and, due to
cancellations in the quark sector, lead to a small mixing
angle between the second- and third-generation quarks.
Some examples of these operators that can lead to the desired hierarchies
between the lepton and quark matrix elements are shown
in Table 2.

\begin{table}
\begin{center}
\begin{tabular}{|c||c|}
\hline
Low-Order Operators & Clebsch-Gordan Factors \\
\hline \hline
& \\
${\bf 16} ~{\bf 45}_X ~{\bf 10} ~{{\bf 45}_Y \over {\bf 45}_X} ~{\bf 16} 
~+
~ {\bf 16} ~{{\bf 45}_Y \over {\bf 45}_X} ~{\bf 10} ~{\bf 45}_X ~{\bf 16}$
& 
$x_u = -3, ~x_d = -\frac{11}{3}, ~x_{\ell} = -17, ~x_{\nu} = ~~5$ \\
& \\
${\bf 16} ~{{\bf 45}_{B-L} \over {\bf 45}_X} ~{{\bf 45}_{B-L} \over
{\bf 45}_X}
 ~{\bf 10} ~{\bf 16} ~+~ 
{\bf 16} ~{\bf 10} ~{{\bf 45}_{B-L} \over {\bf 45}_X} ~{{\bf 45}_{B-L}
\over {\bf 45}_X} ~{\bf 16}$ &
$x_u = ~~2, ~x_d = ~~\frac{10}{9}, ~x_{\ell} = 10, ~x_{\nu} = ~~\frac{34}{25}$ \\
& \\
\hline
\end{tabular}
\end{center}
\caption{\it Low-order operators that, due to
cancellations in the quark sector, 
lead to acceptable range of values for the
$V_{CKM}^{23} / V_{MNS}^{23}$ ratio,
where $V_{MNS}$ is defined in (\ref{vmns}),
despite having smaller differences between the individual
quark and lepton Clebsch-Gordan coefficients.}
\end{table}

Moreover, we also observe the following:

$\bullet$ There exist higher-order operators
that can be less suppressed than others,
under certain conditions. For example,
consider terms of the form
\[
{\bf 16} ~{\bf 45}_i ~{\bf 10} ~\frac{{\bf 45}_j}{{\bf 45}_X}
~\frac{{\bf 45}_k}{{\bf 45}_X} ~{\bf 10} +
{\rm symmetrizing~term}, 
\]
where $i,j,k$ can be any of the
$Y,B-L$ and $T_{3R}$ directions.
If the vev's along the 
$Y,B-L$, $T_{3R}$  and $X$ directions
are not very different, these terms can be 
relatively unsuppressed.
We tabulate some such operators that lead to 
significantly lepton mixing significanmtly larger 
than the quark mixing in Table 3.
We observe that, in these cases,
the lepton mixing tends to be dominated by
the charged-lepton contribution. This is
due to the large $X$ Clebsch-Gordan coefficient of the right-handed
neutrinos, as compared to the charged leptons,
which suppresses the contributions once we
start dividing by powers of ${\bf 45}_X$. The same
suppression occurs for the down-quark contributions,
for this particular type of term. 

\begin{table}
\begin{center}
\begin{tabular}{|c||c|}
\hline
Symmetric Higher-Order Operators & Clebsch-Gordan Factors \\
\hline \hline
& \\
${\bf 16} ~{\bf 45}_Y ~{\bf 10} ~{{\bf 45}_Y \over {\bf 45}_X}
~{{\bf 45}_{B-L} \over {\bf 45}_X}
 ~{\bf 16} ~+~{\rm sym. con.}$ & 
$x_u = 0, ~x_d = \frac{16}{9}, ~x_{\ell} = -48, ~x_{\nu} = 0$ \\
&  \\
${\bf 16} ~{\bf 45}_{B-L} ~10 ~{{\bf 45}_Y \over {\bf 45}_X}
~{{\bf 45}_{B-L} \over {\bf 45}_X}
 ~{\bf 16} ~+~{\rm sym. con.}$ & 
$x_u = 3, ~x_d = -\frac{11}{9}, ~x_{\ell} = -51, ~x_{\nu} = 3$ \\
&  \\
${\bf 16} ~{\bf 45}_{B-L} ~{\bf 10} ~{{\bf 45}_{B-L} \over {\bf 45}_X}
~{{\bf 45}_{{B-L}} \over {\bf 45}_X}
 ~{\bf 16} ~+~{\rm sym. con.}$ & 
$x_u = 0, ~x_d = -\frac{8}{9}, ~x_{\ell} = -24, ~x_{\nu} = -\frac{48}{25}$ \\
&  \\
\hline
\end{tabular}
\end{center}
\caption{\it Illustrative symmetric higher-order operators leading to 
lepton mass-matrix entries that are much larger than the
corresponding quark ones.}
\end{table}

Finally, one may consider operators that are suppressed by powers of
the unknown fundamental scale, which may be of order of $M_{GUT}$
models inspired by M-theory. Some examples of these are shown in Table 4.

\begin{table}[h]
\begin{center}
\begin{tabular}{|c||c|}
\hline
Higher-Dimensional Operators & Clebsch-Gordan Factors \\
\hline \hline
${\bf 16} ~{\bf 45}_Y ~{\bf 10} ~{\bf 45}_X {\bf 45}_{X} ~{\bf 16} ~+~{\rm
sym.con}$ & 
$x_u = -3, ~x_d = 11, ~x_{\ell} = 51, ~x_{\nu} = -75$ \\
${\bf 16} ~{\bf 45}_{B-L} ~{\bf 10} ~{\bf 45}_X {\bf 45}_{{X}} ~{\bf 16}
~+~{\rm sym.con}$ & 
$x_u = 0, ~x_d = 8, ~x_{\ell} = 24, ~x_{\nu} = -48$ \\
\hline
\end{tabular}
\end{center}
\caption{\it Some illustrative operators that could be
relatively unsuppressed in M-thory inspired models,
which lead to domination by neutrino mixing.}
\end{table}

What about the (22) elements in the mass matrices?
Once we fix the (23) entries so
as to obtain large lepton mixing, the (22)
ones need to fall into the following
pattern. For the up-quark mass matrix,
we need the ratio of the (22) and
(33) elements to be smaller than $1/100$
(we can expect to obtain the charm mass from the
off-diagonal elements in the (23) sector)~\cite{ADH10}.
In  the down-quark mass matrix, 
the (23) element contribution to the strange mass
has to be suppressed in order to be consistent with 
the magnitude of $V_{CKM}^{23,32}$ (in the absence of cancellations),
so $m_{s}$ most probably arises from the (22) element,
implying again a very small ratio
$m_D(22)/m_D(33)(M_{GUT})$. 
Finally, we require certain cancellations between
the various entries in the lepton sector,
in order to obtain the correct 
$m_{\mu}/m_{\tau}$ ratio, and this
leads to values of $m_E(22)/m_E(33)(M_{GUT})$
which are typically much larger than the
corresponding ratios in the quark sector.
To obtain these relations, we can

(a) use  operators with similar structures to
those displayed in Tables 3  and 4, considering
also the  possibility of adding 
extra powers of the corresponding adjoint Higgs insertions 
in order to get the correct relations between the (22) 
and (23) elements, or

(b) go to higher-order 
operators, with different structures from that presented above.
However, a detailed operator analysis
along these lines is beyond the scope of this paper.

Finally, we stress again that, in any multi-scale model like the
one discussed above, the particular operators that appear
will be determined by the underlying symmetry.
For instance, one can combine a 
gauged $U(1)$ family symmetry with 
unification in the form of an extended
vertical gauge symmetry \cite{CL}.
In this case, the mass hierarchies in 
a fermion mass matrix are 
determined by the family symmetry,
whilst the splittings between different charge sectors
of the same family are again controlled by the
Clebsch-Gordan factors. In this case as well, one 
predicts {\em exact} Yukawa unification
for the third family, 
together with Clebsch-Gordan relations 
that may reproduce the fermion data.

\section{Neutrino Mixing and Renormalization-Group Effects}

Neutrino mixing arises in analogy to quark mixing,
via a mismatch between the mass eigenstates of neutrinos
and charged leptons, as seen in (\ref{vmns}).
%
%
%
In this section,
we try to understand in more detail the possible structures of
lepton mass matrices that may account for the
various neutrino deficits~\cite{MG,LLR,ELLN2,guid,chlep}. 
%
We concentrate here on the simplest
case, where the problem is well approximated by a two fermion generation
analysis.
%
To see what structures may appear
at low energies, we have to run the various
couplings down to low energies, taking 
neutrino threshold effects properly into account,
as discussed in the next two subsections.

\subsection{Neutrino Renormalization between $M_{GUT}$ and $M_N$}

To study the question of unification, we shall use 
in the numerical analysis the full
two-loop renormalization group evolution of gauge and
Yukawa couplings. However, in order to understand renormalization effects
between $M_{GUT}$ and $M_N$, 
in the small-$\tan\beta$ regime of a supersymmetric
theory where only the top and the Dirac 
neutrino Yukawa couplings contribute
in a relevant way, it is sufficient as a first
illustrative approximation to study the renormalization-group  
equations for the Yukawa couplings at the one-loop level~\footnote{For
another recent study of the renormalization-group equations for
neutrinos, see~\cite{Hab}.}.
These take the following form in a diagonal basis \cite{VS}:
 \bea
 16\pi^2 \frac{d}{dt} \lambda_t&=
 & \left(
 6 \lambda_t^2  + \lambda_N^2
   - G_U\right)  \lambda_t \nonumber \\
 16\pi^2 \frac{d}{dt} \lambda_N&=& \left(
  4\lambda_N^2  + 3 \lambda_t^2
   - G_N \right) \lambda_N \nonumber   \\
 16\pi^2 \frac{d}{dt} \lambda_b &=
 & \left(\lambda_t^2 - G_D \right) \lambda_b \nonumber \\
 16\pi^2 \frac{d}{dt} \lambda_{\tau}&=&\left( \lambda_N^2
  - G_E \right) \lambda_{\tau}
 \label{eq:rg4}
 \eea
where $\lambda_\alpha: \alpha=t,b, \tau ,N$, represent the
third-generation Dirac Yukawa couplings for the up and down quarks,
charged
lepton and neutrinos, respectively, 
and the $G_{\alpha} \equiv \sum_{i=1}^3c_{\alpha}^ig_i(t)^2$ are
functions that depend on the  gauge couplings, with the
coefficients $c_{\alpha}^i$ given in~\cite{VS}. In terms of
the various Yukawa couplings 
$\lambda_{t_0}$, $\lambda_{N_0}$,$\lambda_{b_0},{\lambda_{\tau_0}}$,
at the  unification scale, we have
 \bea
 \lambda_t(t)&=&\gamma_U(t)\lambda_{t_0}\xi_t^6\xi_N ~~~~~ 
\lambda_N(t)=\gamma_N(t)\lambda_{t_0}\xi_t^3\xi_N^4\\
 \lambda_b(t)&=&\gamma_D(t)\lambda_{b_0}\xi_t ~~~~~~~~~ 
\lambda_{\tau}(t)=\gamma_E(t)\lambda_{\tau_0}\xi_N
\\
 \gamma_\alpha(t)&=&  \exp\left ({\frac{1}{16\pi^2}\int_{t_0}^t
  G_\alpha(t) \,dt} \right ) = \prod_{j=1}^3 \left( 
\frac{\alpha_{j,0}}{\alpha_j}
 \right)^{c_\alpha^j/2b_j} \\
 \xi_i&=& \exp
\left ({\frac{1}{16\pi^2}\int_{t_0}^t \lambda^2_{i}dt} \right )
 \eea
It is obvious that the ratio of bottom and tau Yukawa couplings at any
scale depends on the integral~\cite{LLR}:
\begin{equation}
\xi_N = \exp
\left ({\frac{1}{16\pi^2}\int_{t_0}^t \lambda^2_{N}dt} \right )
\label{xiN}
\end{equation}
In the absence of neutrino couplings, this factor is equal
to one.
However, when $\lambda_{N_0} \ne 0$, $\xi_N$  becomes lower
than one and
affects the unification conditions. The quantity (\ref{xiN})
therefore plays a key role in our subsequent analysis.

At this stage, we need to remember that
the $b$--$\tau$ equality at the GUT scale refers to the
$(3,3)$ entries of the charged-lepton and
down-quark mass matrices, whereas the detailed structure of the
mass matrices may not be predicted by the GUT without some
extra assumption(s), as discussed in the previous and 
subsequent sections. Therefore,
it is relevant to assume mass textures
with the property that the $(m^{diag}_E)_{33}$ and $(m^{diag}_D)_{33}$
entries are no longer equal 
after diagonalization at the GUT scale~\cite{LLR}.
To understand the effect, we consider
a $2 \times 2$ example, and  assume that the off-diagonal terms
in the down-quark mass matrix $m_D$ are small compared to
the (33) element, whereas this is not the case for the
charged-lepton mass matrix. In this case, one can approximate the 
down-quark and charged-lepton mass matrices at the GUT scale by
\bea
\label{d0e0}
m_{D}^0 = A 
 \left (
 \begin{array}{cc}
c & 0 \\
0 & 1 
 \end{array}
 \right), ~~~ m_{E}^0 = A 
 \left (
 \begin{array}{cc}
x^2 & x \\
x & 1 
 \end{array}
 \right),
\eea
where $A$ may be identified with $m_b(M_{GUT})$, the 
bottom quark mass at the scale $M_{GUT}$.
These textures ensure that 
 $(m_{D}^0)_{33}= (m_{E}^0)_{33}$ at the GUT scale. Moreover,
the form of $m_E^0$ is such that the hierarchical relation
between the two mass eigenvalues is obtained, $m_3 \gg m_2$. 
Note that, in this simple example, we work with textures
that are symmetric before renormalization, which
affects differently the left- and right-handed states.
At low energies, the eigenmasses are obtained by
diagonalising the renormalized Yukawa matrices, which is equivalent to
diagonalising the quark and charged-lepton Yukawa matrices
at the GUT scale, and then evolving the eigenstates and
the mixing angles separately. In this way,
we see that the trace of the charged-lepton mass matrix,
which gives the higher eigenvalue,
is not 1 but $1+x^2$, and therefore 
the effective $\lambda_b$ and $\lambda_\tau$ 
are not equal after diagonalization.

To simplify our analysis, we assume  that 
$M_{\nu_R}$ is diagonal with degenerate eigenvalues.
For hierarchical neutrino masses, however, the
results can be generalized for arbitrary heavy
Majorana neutrino masses by using the results of 
Appendix B. 
Since we are working in the small-$\tan\beta$ regime,
in which the renormalization-group 
effects of the charged-lepton
Yukawa couplings can be neglected, 
it is easier to pass to a basis in which
the Dirac neutrino-mass matrix is diagonal, by performing
an SU(2)-invariant rotation of the Dirac mass matrices
for charged and neutral leptons. In this basis,
the renormalization-group effect of the potentially large neutrino
Yukawa couplings in the run from $M_{GUT}$ down to $M_N$ is simply
parametrized by the integral $\xi_N$.

Let us also assume that the initial neutrino texture at $M_{GUT}$ is
\bea
\label{initial}
m_\nu^D = B 
 \left (
 \begin{array}{cc}
y^2 & y \\
y & 1 
 \end{array}
 \right),
\eea
leading to
\bea
 m_{eff}^0 =   m_{\nu}^{D}\cdot M_{\nu_R}^{-1} \cdot
m_{\nu}^{D^T} = B^2 (y^2+1) 
 \left (
 \begin{array}{cc}
y^2 & y \\
y & 1 
 \end{array}
 \right)
\eea 
It should be stressed that we
have, in general, a relative phase between the (12) elements
of the  mass matrices $m_{\nu}^D$ and $m_E^0$.
This phase should in principle be
taken into account, since it determines whether the
mixing 
between the charged-lepton and the neutrino
sectors, is {constructive} or
{destructive}~\cite{MG}. Below, we take the mass 
matrix elements to be real numbers, so we need only consider 
the distinct cases $x y > 0$ and $x y < 0$.
Since the matrix diagonalising 
$m_{eff}$ is 
\bea
V_{\nu} =
\frac{1}{\sqrt{1+y^2}}
 \left (
 \begin{array}{cc}
1 & -y \\
y & 1 
 \end{array}
 \right),~~~
\eea 
the form of $m_E$ at the GUT
scale is, in the basis where the neutrinos
are diagonal, 
\bea
\label{mlepmg}
 m_{E}(M_{GUT}) =
  A
\left (
 \begin{array}{cc}
a & \sqrt{ab} \\
\sqrt{ab} & b
 \end{array}
 \right)
\eea
where
\begin{equation}
a \equiv (x-y)^2/(1 + y^2), \;\;\; b \equiv (xy+1)^2/(1 + y^2).
\label{aandb}
\end{equation}
Taking into account the running
of the charged-lepton mass matrix, at the scale $M_N$
one has:
\bea
\label{mlepmn}
 m_{E}(M_N)
= \tilde{A}  \left (
 \begin{array}{cc}
a & \sqrt{ab} \\
\xi_N \sqrt{ab} & \xi_N b
 \end{array}
 \right),
\eea
where $\tilde{A}$ is a coefficient that
contains flavour-independent renormalization-group effects, 
and $\xi_N$, (\ref{xiN}) is
the integral associated with
the running of the neutrino Yukawa coupling.
Since $m_{E}(M_N)$ is non-symmetric, the
left- and right-handed lepton mixing angles are different. The
left-handed mixing angle at $M_N$ is calculated by
diagonalising the hermitian matrix
$m_{E}(M_N) m_{E}^{\dagger}(M_N)$, and is given by
\bea
\sin2\theta_{23} (M_N) = \frac{2 \sqrt{ab}  \xi_N}
{a \; + \; \xi_N^2 \;b }
\label{lmixing}
\eea

\subsection{Neutrino Renormalization below $M_N$}

Before proceeding with the numerical analysis,
for completeness we first discuss renormalization
below the right-handed Majorana mass scale. Here,
${\lambda}_N$ decouples and the relevant running
is that of the effective neutrino mass operator:
\bea
    8\pi^2 {d\over dt}{ \k} & = & \left[-({3\over 5} g_1^2+3 g_2^2)+
3 \lambda_t^2 \right] \k 
\label{MASSES}
\eea
This implies that an initial texture 
$m_{eff}(M_N)^{ij}$
at $M_N$ becomes at  a lower scale
\bea
m_{eff}  \propto
I_g \cdot I_t \cdot  m_{eff}(M_N)
\eea
where we have not written  explicitly the dependence on 
$I_i =   exp[\frac 1{8\pi^2}\int_{t_0}^t \lambda_{i}^2 dt]$, with the 
subindex $i$ referring to the 
charged-lepton flavours $e, \mu$ and $\tau$, and where
\bea
 I_g& =& exp\left[\frac{1}{8\pi^2}\int_{t_0}^t(-c_i g_i^2 dt)\right]\\
I_t &=& exp\left[ \frac{3}{8\pi^2}\int_{t_0}^t  \lambda_t^2 dt \right].
\eea
The running of 
the lepton  mixing angle $\theta_{23}$
is given by \cite{Bab,run}
\bea
    16\pi^2 {d\over dt}{\sin^2 2\theta_{23}} & =& -2\sin^2 2\theta_{23}
        (1-\sin^2 2\theta_{23}) \nonumber 
  (\lambda_{\tau}^2-\lambda_{\mu}^2){\k^{33}+\k^{22}\over
\k^{33}-\k^{22}}.  
\label{MIXING}
\eea
We see from (\ref{MIXING})
that, for small $\tan\beta$, the 
renormalization effects
on the mixing are small below $M_N$, reflecting the stability
of the texture.
Thus, in practice, the mixing  freezes at the scale $M_N$ and the angle
defined in 
(\ref{lmixing}) is the one we compare with
the low-energy data on mixing in the lepton sector.
However, the exact values of the neutrino masses 
depend on $I_g$ and $I_t$. 

\subsection{Conditions for Maximal Mixing}

The best fit to the Super-Kamiokande data is obtained for
a value of the mixing angle in the lepton sector $\sin 2 \theta_{23}$
close to unity: $\sin^2 2 \theta_{23} > 0.9$. 
We see from (\ref{lmixing}) that the requirement 
of maximal mixing, $\sin 2 \theta_{23} = 1$,  is equivalent to the condition
\begin{equation}
\sqrt{a} = \sqrt{b} \; \xi_N.
\end{equation}
Using the expressions (\ref{aandb}) for $a$ and $b$, we
rewrite this expression as
\begin{equation}
x - y = \pm  \; \xi_N \; ( x y + 1).
\end{equation}
There are two solutions for this equation, associated with the
two signs of the square roots, which should be chosen in such a way
that $|x|$ and $|y|$ are less than one. Hence
the requirement of maximal mixing, $\sin 2 \theta_{23} \simeq 1$, 
is equivalent to the relations
\begin{eqnarray}
\label{ymaxmix}
y & = & \frac{ x - \xi_N}{ 1 + \xi_N \; x} \;\;\;\;\;\;\;\;\;\;\;\;\
{\rm for} \;\;\; x  \geq 0 
\nonumber\\
y & = & \frac{ x + \xi_N}{ 1 - \xi_N \; x} \;\;\;\;\;\;\;\;\;\;\;\;\
{\rm for} \;\;\; x  \leq 0. 
\end{eqnarray}
Therefore, the requirement of maximal mixing at low energies gives
a simple relation between the entries of the 
neutrino and charged-lepton mass matrices at high energies, as
a function of the neutrino renormalization-group factor $\xi_N$, which,
as we shall show, is bounded by $1 > \xi_N > 0.8$. 

\section{Numerical Analysis}

We use two different codes for our numerical analysis,
one taking a bottom-up and the other a top-down
approach to gauge- and
Yukawa-coupling unification~\footnote{As already remarked, we have
checked that the inclusion of neutrinos in the
renormalization-group equations for the gauge couplings does not greatly
affect the conditions for gauge unification.}. 
We use the two-loop renormalization-group evolution of 
the gauge and Yukawa couplings in both programs,
except for the running of $m_{eff}$,
where we use one-loop formulae.
In the range between $M_{GUT}$ and $M_N$, we
take into account the running of $\lambda_N$, and run
with the supersymmetric $\beta$ functions.
At the scale $M_N$, the effective lepton mixing freezes, as
mentioned earlier,
but the evolution of $m_{eff}$ still has to be taken into account,
since it changes the
magnitudes of the neutrino mass eigenvalues.
Thus, between $M_N$ and $M_{SUSY}$ we include
the running of $m_{eff}$ and continue to run
with the
supersymmetric $\beta$ functions. 
Below $M_{SUSY}$, we
run with the Standard Model $\beta$ functions, and
continue to include the running of $m_{eff}$,
which is now described by the equation
\begin{equation}
16 \pi^2 {d m_{eff} \over d  t}
= ( - 3 g_2^2 + 2 \lambda + 2 S ) m_{eff}
\end{equation}
where $\lambda$ is the Higgs coupling:
$M_H^2 = \lambda v^2$,
and $ S \equiv 3 \lambda_t^2$~\cite{Bab}.  
Finally, for the running between $M_Z$ and $m_b$, we take
into account the decoupling of both $\lambda_t$ and $g_2$
from the running of the couplings.

As a basis for our numerical studies, we first consider
evolution of the (33) matrix elements as if these
were equal to the heav`ier eigenvalues of the mass matrices,
i.e., ignoring mixing effects in the runs. 
Furthermore, we do not demand precise unification of couplings,
but we choose a scale $M_{GUT} \simeq 1.5 \; 10^{16}$ GeV,
for which approximate unification of the three gauge couplings
holds at the few per-mille level for all the cases analysed in
this article. We choose a characteristic soft supersymmetry
breaking scale of the order of 1 TeV, at which we decouple all
supersymmetric particles, and for which approximate unification
of gauge couplings 
can be achieved for values of $\alpha_3(M_Z) \simeq 0.118$~\cite{CPW}.
We then plot the ratio $m_{\tau}/m_b(M_{GUT})$, as a
function of $M_N$, for fixed values of the
light neutrino mass $m_{\nu}$ and of $\tan\beta$. 
This is shown in Figs.~1, 2 and 3 for the neutrino 
mass values $m_{\nu} = 0.03$, 0.1 and 1 eV, respectively. 
In each figure, the lines are truncated when the value of $M_N$ is
such that the neutrino Yukawa coupling enters the non-perturbative
regime at scales below $M_{GUT}: \lambda_N^2(M_N)/4\pi 
\simeq 1$~\footnote{In this analysis, 
we have ignored the  threshold corrections
to the fermion Yukawa couplings induced by the squark and slepton
mixing parameters, since these corrections are 
strongly model dependent and small at these values of $\tan\beta$, and
therefore they do not affect the overall picture displayed 
in Figs.~1, 2 and 3~\cite{HEC,CCPW}.}

\begin{figure}[t]  
\centerline{
\psfig{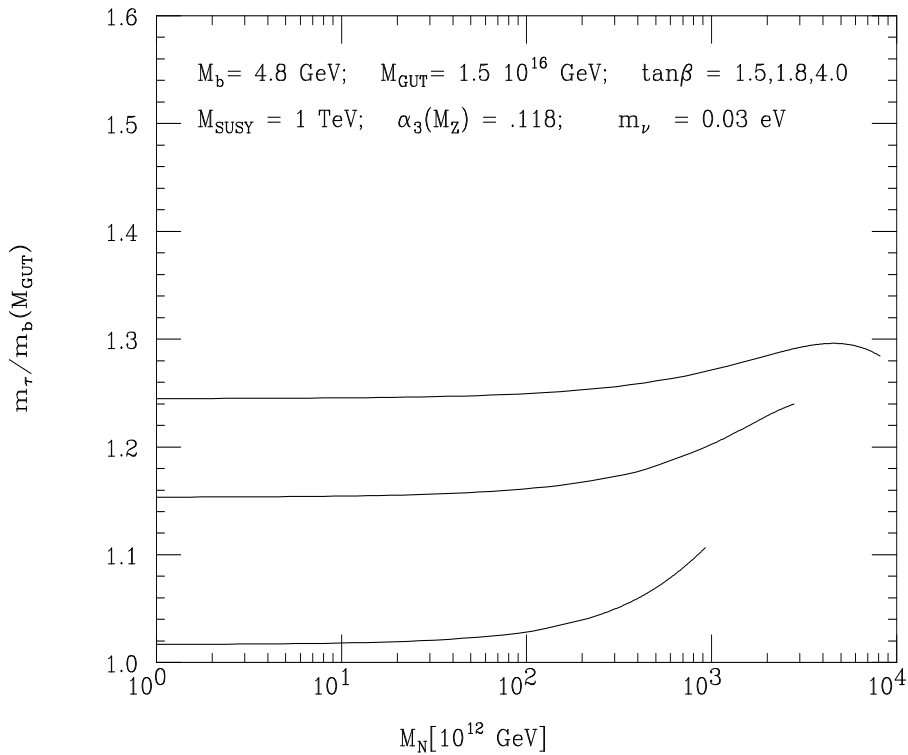}}
\caption
{\it The ratio
$m_\tau/m_b(M_{GUT})$ as a function of
$M_N$, with the choice  
$m_{\nu}$ = 0.03 eV and
for different values of 
$\tan\beta$: from bottom to top, $\tan\beta$ = 1.5, 1.8 and 4, respectively. }
\end{figure}

\begin{figure}[t]  
\centerline{
\psfig{figure=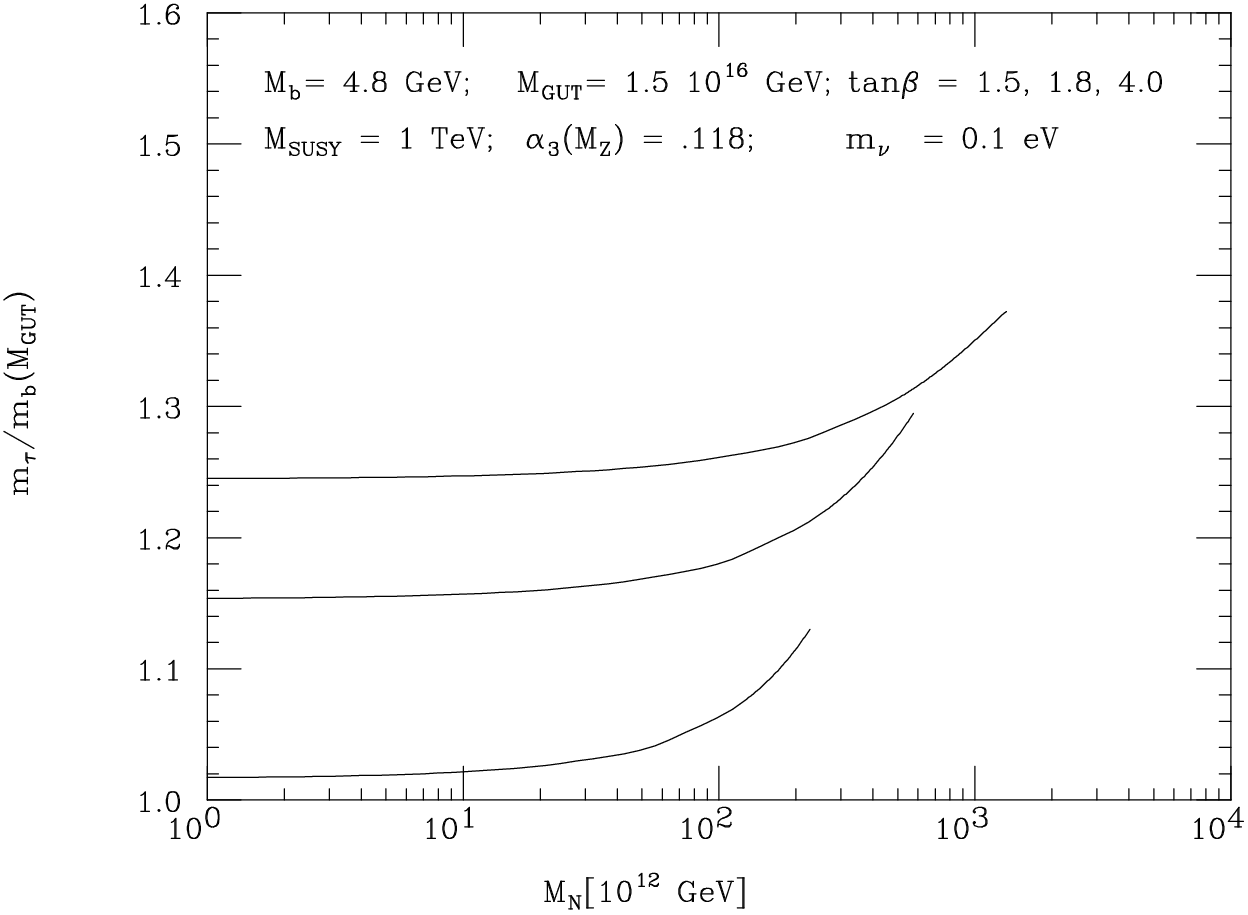,height=12cm,width=10cm,angle=90}}
\caption
{\it The ratio
$m_\tau/m_b(M_{GUT})$ as a function of
$M_N$, with the choice  
$m_{\nu}$ = 0.1 eV, for the same values of $\tan\beta$, from bottom
to top,  as
in Fig. 1.
}
\end{figure}

\begin{figure}[t]  
\centerline{
\psfig{figure=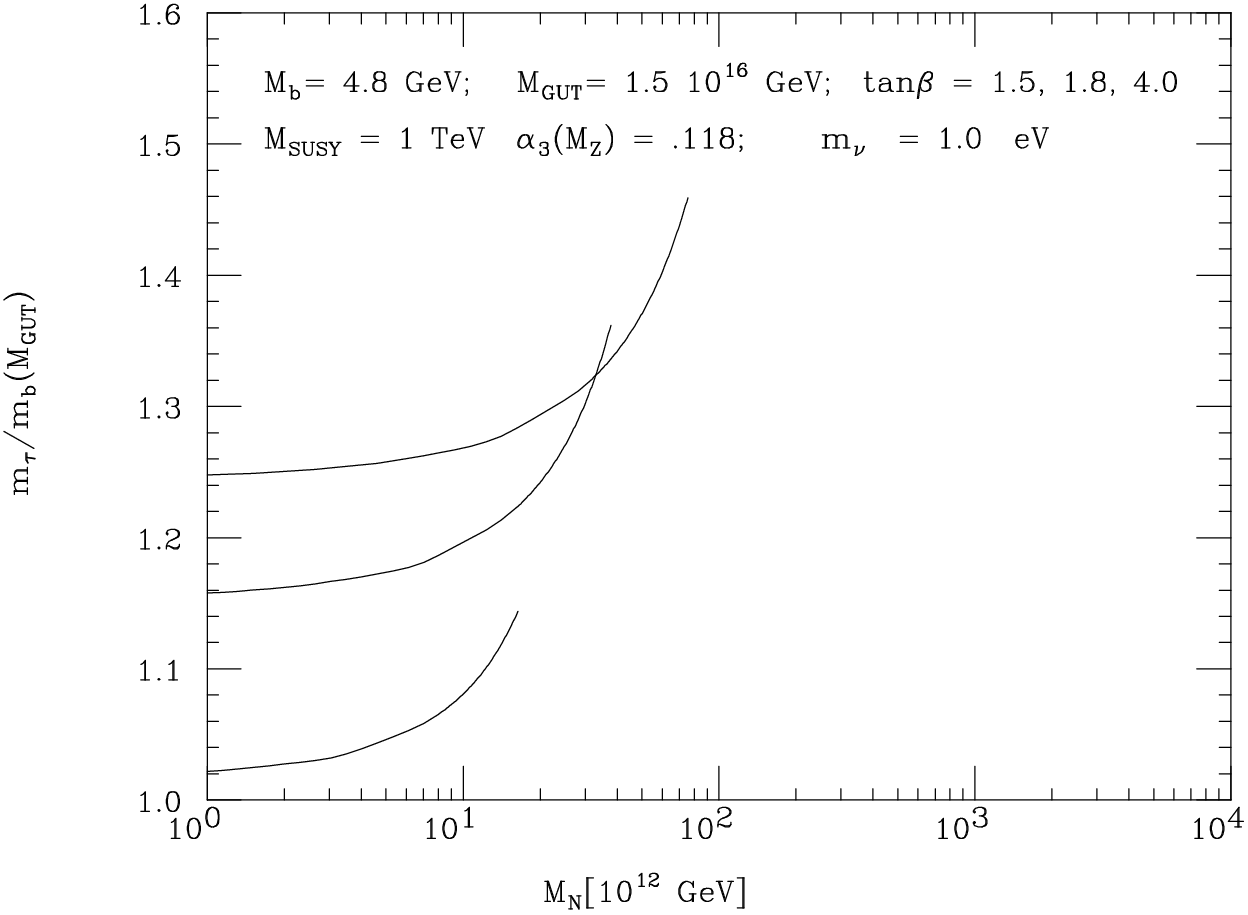,height=12cm,width=10cm,angle=90}}
\caption
{\it The ratio
$m_\tau/m_b(M_{GUT})$ as a function of
$M_N$, with the choice  
$m_{\nu}$ = 1 eV, for the same values of $\tan\beta$, from bottom
to top, as in Fig. 1.
}
\end{figure}

Later we will reinterpret the results of the figures taking into
account the mixing effects in the running,
 but already at this stage we can make some 
interesting observations:

(i) For small $\lambda_N$, corresponding in the see-saw model to
small $M_N$, the appearance of
the neutrino masses does not play a major role. Therefore,
for small values of $\tan\beta$
in the region of the top-quark mass infrared fixed-point solution, we 
obtain $b-\tau$ unification, from which we see a deviation
as $\tan\beta$ increases.

(ii) As $\lambda_N$ becomes larger for fixed $\tan\beta$, 
corresponding to a larger $M_N$ in the see-saw model,
the neutrino coupling lowers $\lambda_{\tau}$ with
respect to $\lambda_b$. Therefore, in order
to obtain the correct value of $m_b/m_\tau$
at low energies, we need to
start with lower values of $\lambda_b/\lambda_\tau(M_{GUT})$.

(iii) As the coupling $\lambda_N$ increases, we expect
that at some stage the corresponding value of $M_N$ gets
so close to the GUT scale that, after reaching a peak,
the effects on $b-\tau$ unification decrease again,
because of their dependence on $\ln(M_N/M_{GUT})$.
This explains the behaviour of $m_{\tau}/m_b(M_{GUT})$ for 
the choice $\tan\beta = 4$ in Fig.~1.
For the other values of $\tan\beta$ displayed in Fig.~1, and also for
all values of $\tan\beta$ displayed in Figs.~2 and 3, the Dirac neutrino 
coupling is so large
that $\lambda_N$ enters the non-perturbative regime before
this peak is reached.

(iv) As stated above, in Fig.~3 we present, for completeness,  
the results on $m_{\tau}/m_b(M_{GUT})$
for a neutrino mass $m_{\nu} = 1$ eV. Such a value of the neutrino mass
is not compatible with the assumption of hierarchical neutrino masses
which we favour in this paper. However, Fig.~3 serves to document
the important constraints on parameter space that are
imposed in this case by requiring that one avoids the
non-perturbative regime. We do
not analyze this value of $m_{\nu}$ any further in this paper.

%
%

\subsection{Neutrino Mixing Effects}

The main effects of the neutrino Yukawa couplings on the running of the
lepton Yukawa coupling matrix elements have been presented 
in (\ref{mlepmn}). In the above, when plotting the ratio
of the bottom to tau Yukawa couplings, we worked in the
basis in which the neutrinos are diagonal, and ignored the mixing
in the charged-lepton sector. We now discuss how to 
interpret the precise unification relation of the (33) elements in the
down-quark and charged-lepton sectors in view of the results
shown in Figs.~1 and 2.

In the presence of lepton mixing, and in the basis
in which the neutrino Dirac mass matrix is diagonal, 
the quantity that is renormalized
in the way displayed in the Figures as the tau Yukawa
coupling is in fact the (33) element of the charged-lepton mass matrix.
Hence, if we denote by $\tilde{m}_{\tau}$ the extrapolated value of
the $\tau$ mass shown in the Figures, the value of the
(33) element at the GUT scale is given in the presence of mixing by,
\begin{equation}
(m_E)_{33}(M_{GUT}) = \tilde{m}_{\tau}(M_{GUT})  \times
\frac{(m_E)_{33}}{m_{\tau}}
\end{equation}
or, equivalently,
\begin{equation}
(m_E)_{33}(M_{GUT}) = \tilde{m}_{\tau}(M_{GUT}) \times
\frac{\xi_N \; b}{\sqrt{(a + b) ( a + b \xi_N^2)}}
\end{equation}
Since $(m_E)_{33}(M_{GUT}) = A \; b$ in this basis, we find
\begin{equation}
\frac{\tilde{m}_{\tau}(M_{GUT})}{A} = 
\frac{\sqrt{ (a + b)(a +\xi_N^2 b)}}{\xi_N}.
\label{bottomt}
\end{equation}
Hence, in the presence of mixing
in the lepton sector, we {\it expect} a mismatch of the 
extrapolated value of the $\tau$ Yukawa coupling, as given by
the above expression.
 
The factor $\xi_N$ can be obtained simply
from the Figures, by observing that the difference between the
extrapolated
values of $\tilde{m}_{\tau}$ for vanishing 
and non-vanishing neutrino Yukawa couplings
is given, at one-loop, 
by $1/\xi_N$. Hence, for the same value of $\tan\beta$,
one should take the obtained value of $\tilde{m}_{\tau}/m_b$
for a given value of $M_N$ and divide it by its value at low
values of $M_N \simlt 10^{12}$ GeV to obtain, in a very
good approximation, the desired factor $1/\xi_N$.

\subsection{$b-\tau$ Unification and Maximal Mixing}

We now combine the above information with the condition of 
maximal mixing. Substituting the value of $y$ required 
for maximal mixing, (\ref{ymaxmix}), one obtains, 
\begin{eqnarray}
a & = &  \frac{ \xi_N^2 \; ( x^2 + 1)}{\xi_N^2 + 1}
\nonumber\\
b & = & \frac{ 1 + x^2 }{\xi_N^2 + 1}.
\end{eqnarray}
Further substituting these expressions into (\ref{bottomt}), one obtains
\begin{equation}
\label{comparison}
\frac{\tilde{m}_{\tau}(M_{GUT})}{m_b(M_{GUT})} 
= (1 + x^2) \sqrt{\frac{2}{1 + \xi_N^2}},
\end{equation}
where we have assumed that unification takes place
as described in (\ref{d0e0}): $m_b(M_{GUT}) = A$.
Since $0.8 > \xi_N > 1$, the ratio $\tilde{m}_{\tau}/m_b$ at
$M_{GUT}$ given by (\ref{comparison}) must be close to  $(1 + x^2)$.

We may now address the question of unification. We use
the results of Fig.~1 to specify the value of $x$ 
for which one obtains the
desired unification relation. For this, one has first to read the
value of $\xi_N$ from the figure, as explained before, and then compare
the obtained value of $\tilde{m}_{\tau}/m_b$ with (\ref{comparison}).
Since all the values of $\tilde{m}_{\tau}(M_{GUT})/m_b(M_{GUT})$
encountered in Fig.~1 are lower than 1.5, a solution 
for $x$ consistent with unification can always be obtained. With this
solution at hand, one can determine from (\ref{ymaxmix}) the
value of $y$ needed to obtain maximal mixing.

For instance, 
for low values of $\tan\beta$ close
to the fixed point, and low values of $M_N \simeq 10^{12}$ GeV,
we find that 
$\xi_N \simeq 1$ and $b-\tau$ mass unification demands $x \simeq 0$.
Thus, all the mixing should be located in the neutrino sector
in the original basis at the scale $M_{GUT}$. 
On the contrary, for the same value of
$\tan\beta$, but larger values of $M_N$ ($1/\xi_N$) for which unification
of neutrino and top Yukawa coupling can take place, 
we find that $x \simeq 0.25$--0.3
and therefore $y \simeq -(0.5$--0.55), so moderate mixing in both the lepton
Dirac mass matrices is required at the GUT
scale.

\begin{table}[tbp]
\begin{center}
\begin{tabular}{|c|c|c|c|c|c|c|c|c|}
\hline
$M_N[10^{13}$ GeV] & 1 & 10 & 20 & 50 & 70 & 150 & 250 & 400 \cr
\hline
$\tan\beta = 1.5$ &0.13 &0.15 & 0.17 & 0.21 &0.23 & & &\cr
\hline
$\tan\beta = 1.8$  &0.39 &0.40 &0.40&0.41&0.42 &0.43 & 0.44&\cr
\hline
$\tan\beta = 4.0$&0.50 &0.50& 0.50&0.50&0.50 &0.51 &0.52 & 0.52\cr
\hline
\end{tabular}
\end{center}
\caption{\it Values of $x$ leading to $b-\tau$
Yukawa coupling unification for $m_{\nu}= 0.03$ eV,
for different choices of $\tan\beta$ and $M_N$.}
\label{tab:x003}
\end{table}
\begin{table}[tbp]
\begin{center}
\begin{tabular}{|c|c|c|c|c|c|c|c|c|}
\hline
$M_N[10^{13}$ GeV] & 1 & 10 & 20 & 50 & 70 & 150 & 250 & 400 \cr
\hline
$\tan\beta = 1.5$ &-0.77 &-0.73 & -0.69 &-0.62& -0.58 & & &\cr
\hline
$\tan\beta = 1.8$  &-0.44 &-0.43&-0.42&-0.40&-0.39 &-0.37 &-0.35 &\cr
\hline
$\tan\beta = 4.0$&-0.34 &-0.33&-0.33&-0.32&-0.32 &-0.31 &-0.30 &-0.29\cr
\hline
\end{tabular}
\end{center}
\caption{\it Values of $y$ 
leading to $b-\tau$ Yukawa coupling
unification for $m_{\nu}= 0.03$ eV,
for different choices of $\tan\beta$ and $M_N$.}
\label{tab:y003}
\end{table}

\begin{table}[tbp]
\begin{center}
\begin{tabular}{|c|c|c|c|c|c|c|c|c|}
\hline
$M_N[10^{12}$ GeV ]& 1 & 10 & 50 & 80 & 100 & 200 & 400 & 800 \cr
\hline
$\tan\beta = 1.49$ &0.13 &0.14 & 0.17 & 0.18 &0.20 & 0.26& &\cr
\hline
$\tan\beta = 1.8$  &0.39 &0.39 &0.40&0.41&0.41 &0.43 & 0.45&\cr
\hline
$\tan\beta = 4.0$&0.50 &0.50& 0.50&0.50&0.50 &0.51 &0.52 & 0.54\cr
\hline
\end{tabular}
\end{center}
\caption{\it Values of $x$ leading to $b-\tau$ Yukawa coupling
unification for $m_{\nu}= 0.1$ eV,
for different choices of $\tan\beta$ and $M_N$.}
\label{tab:x01}
\end{table}
\begin{table}[tbp]
\begin{center}
\begin{tabular}{|c|c|c|c|c|c|c|c|c|}
\hline
$M_N[10^{12}$ GeV] & 1 & 10 & 50 & 80 & 100 & 200 & 400 & 800 \cr
\hline
$\tan\beta = 1.49$ &-0.76 &-0.75 & -0.70 &-0.66& -0.62 &-0.52 & &\cr
\hline
$\tan\beta = 1.8$  &-0.44 &-0.43&-0.42&-0.41&-0.40 &-0.37 &-0.33 &\cr
\hline
$\tan\beta = 4.0$&-0.34 &-0.34&-0.33&-0.33&-0.32 &-0.31 &-0.29 &-0.26\cr
\hline
\end{tabular}
\end{center}
\caption{\it Values of $y$ 
leading to $b-\tau$ Yukawa coupling unification for $m_{\nu}= 0.1$ eV,
for different choices of $\tan\beta$ and $M_N$.}
\label{tab:y01}
\end{table}

We summarize in Tables \ref{tab:x003} and \ref{tab:y003} 
the values of $x$ and
$y$ needed to obtain mass unification for a neutrino mass of order 0.03
eV.
The same is done in Tables \ref{tab:x01} and \ref{tab:y01} for a 
neutrino mass of order 0.1 eV. We observe that the range of values of
$x$ and $y$ needed to achieve unification depends strongly
on $\tan\beta$, but does not depend much on the exact value of
the neutrino mass. For a given value of $\tan\beta$, 
the values of $x$ and $y$
depend, in a first approximation, only on the product $m_{\nu} \times
M_N$~\footnote{Although we do not discuss this case in detail, we
comment that, at large $\tan\beta$, bottom-$\tau$ unification is
consistent with neutrino masses, even in the absence of lepton mixing,
as a result of the bottom Yukawa coupling effects
and the corrections from sparticle loops to 
$m_b$ (for a discussion of these effects see, for example, 
Ref. \cite{btaut}).}.

\begin{figure}[t]  
\centerline{
\psfig{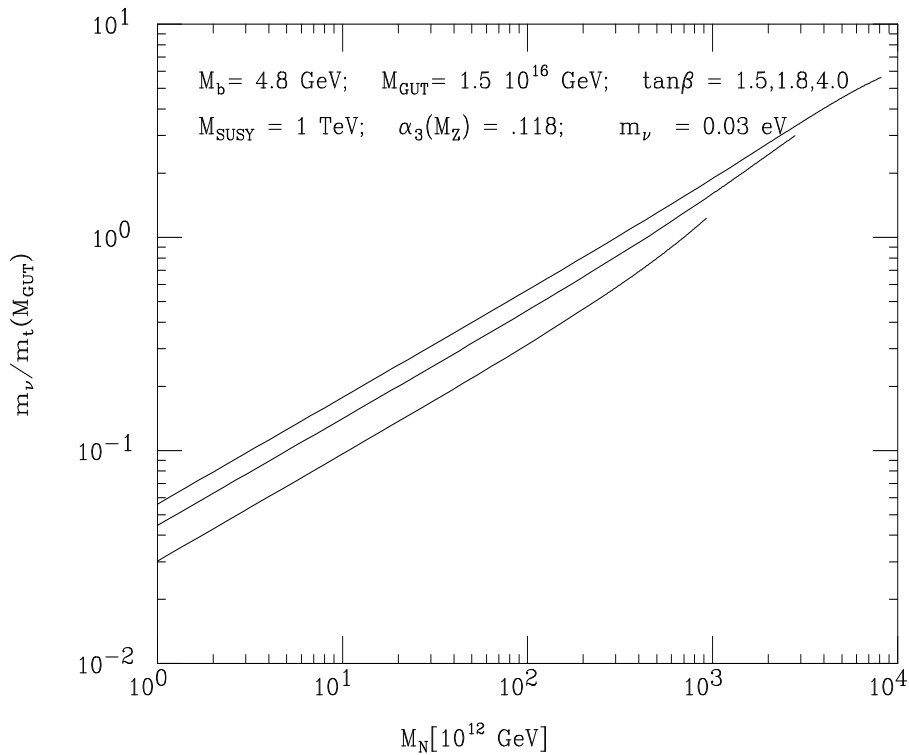}}
\caption
{\it The ratio
$m_\nu/m_t(M_{GUT})$ as a function of
$M_N$, with the choice  
$m_{\nu}$ = 0.03 eV, for the same values of $\tan\beta$, from
bottom to top, as in Fig. 1.
}
\end{figure}
\begin{figure}[t]  
\centerline{
\psfig{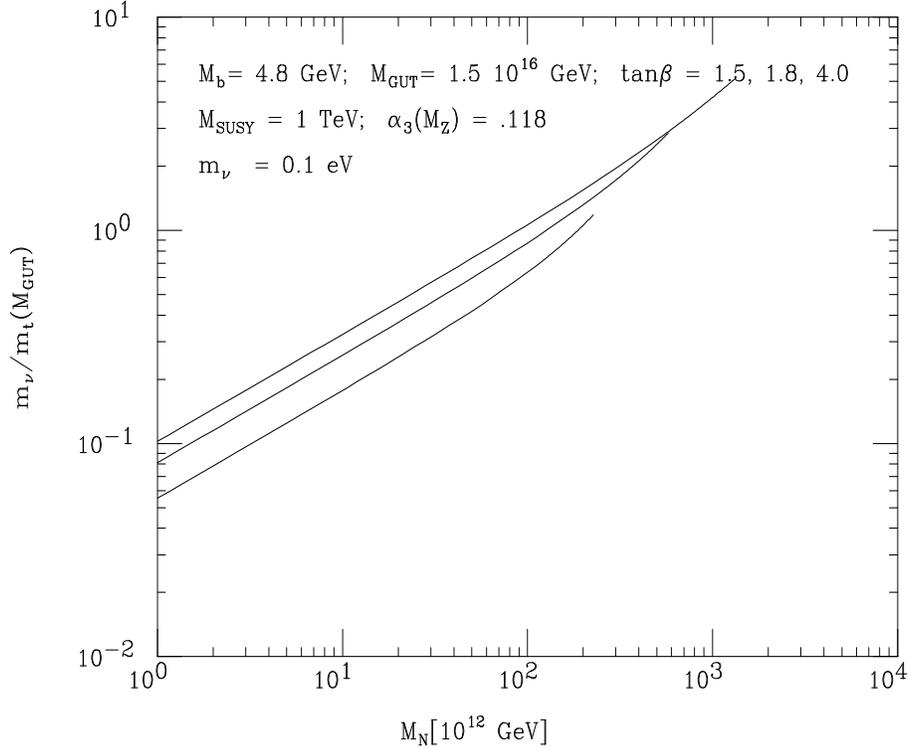}}
\caption
{\it The ratio
$m_\nu/m_t(M_{GUT})$ as a function of
$M_N$, with the choice  
$m_{\nu}$ = 0.1 eV, for the same values of $\tan\beta$, from
bottom to top, as in Fig. 1.
}
\end{figure}

\subsection{On the Unification of $\lambda_{N}$ and $\lambda_t$}

In $SO(10)$ models, in which the third generation Yukawa couplings
appear from the $SO(10)$ invariant Yukawa interaction 
${\bf 16}_3 {\bf 10} {\bf 16}_3$, not only the bottom and Yukawa couplings
unify,
but also the neutrino and the top Yukawa couplings take equal
value at $M_{GUT}$~\cite{VS,Rat}.
It is therefore interesting to see when unification of the neutrino
and the top-quark Yukawa couplings takes place. Since we are working in
the
basis in which the neutrino and top-quark Dirac mass matrices are
diagonal,
unification takes place when the ratio of the neutrino
to the top-quark Yukawa couplings is given by,
\begin{equation}
\frac{\lambda_{N}(M_{GUT})}{\lambda_{t}(M_{GUT})} = 1 + y^2.
\end{equation}
One can now use the values of $y$ displayed in the
Tables \ref{tab:y003} and \ref{tab:y01} to investigate the values
of $M_N$ for which $b-\tau$  and $t-\nu$ unification take place as a
function
of $\tan\beta$ and $M_N$.

Figures 4 and 5 show the behaviour of the ratio of the 
neutrino Yukawa coupling to the top Yukawa coupling at the
scale $M_{GUT}$.
Close to the infrared fixed
point of the top-quark mass: $\tan\beta \simeq 1.5$, $t-\nu$ Yukawa
unification can only take place at large values of $M_N$ for which
the neutrino Yukawa coupling becomes strong, namely, for values
$M_N \simeq 9 \; 10^{14}$ GeV and $M_N \simeq 2.2 \; 10^{14}$ GeV
for $m_{\nu} = 0.03$ eV and $m_{\nu} = 0.1$ eV, respectively. 
For $m_{\nu} = 0.03$ eV and $\tan\beta = 1.8$ ($\tan\beta = 4$)
$t-\nu$ Yukawa unification can be achieved for $M_N \simeq 5 \; 10^{14}$ GeV
($M_N \simeq 3.5 \; 10^{14}$ GeV). 
For $m_{\nu} = 0.1$ eV, instead, and
$\tan\beta = 1.8$ ($\tan\beta = 4$), $t-\nu$ Yukawa unification
can be achieved for $M_N \simeq 1.4 \; 10^{14}$ GeV 
($M_N \simeq 1.0 \; 10^{14}$
GeV).

\section{Comparison with Textures Derived from Exact $SU(5)$ Relations}

In the above, we have analyzed the consequences of assuming quark and
lepton mass matrix textures that are generated via higher-order
operators, and hence violate the exact $SU(5)$ relation between the
left-handed lepton and right-handed down-quark sectors. This relation
could be fulfilled by taking, for instance, the following mass matrix
textures at the GUT scale~\cite{HO,gui1}
\bea
m_{D}^0 = A 
 \left (
 \begin{array}{cc}
0 & 0 \\
x & 1 
 \end{array}
 \right), ~~~ m_{E}^0 = A 
 \left (
 \begin{array}{cc}
0 & x \\
0 & 1 
 \end{array}
 \right).
\eea
We assume that the neutrino mass matrix at the GUT scale is
still symmetric and parametrized by the parameter $y$, as in 
(\ref{initial}). In the basis in which the neutrinos are
diagonal, the lepton mass matrix takes the form
\bea
m_{E}(M_{GUT}) = \frac{A}{1 + y^2} 
 \left (
 \begin{array}{cc}
-y (x-y)  & x - y \\
-y (xy+1) & xy +1 
 \end{array}
 \right).
\eea
and hence, following the same procedure as in the previous case, we
find that the low-energy  form of the lepton mass matrix is given
by 
\bea
m_{E}(M_{N}) = \frac{\tilde{A}}{1 + y^2} 
 \left (
 \begin{array}{cc}
-y (x-y)  & x - y \\
-\xi_N y (xy+1) & \xi_N(xy +1) 
 \end{array}
 \right).
\eea
>From the above,  we obtain the following value of 
the lepton mixing angle $\sin 2 \theta_{23}$:
\bea
\sin 2 \theta_{23} = 2 \xi_N \frac{(x- y)(xy+1)}{\xi_N^2(xy+1)^2 + (x-y)^2}.
\eea
which is exactly the same as in (\ref{lmixing}). Hence the
conditions for maximal mixing are satisfied for exactly the same
relation between $y$ and $x$ as in (\ref{ymaxmix}). However, the relation
between the extrapolated value of $\tilde{m}_{\tau}(M_{GUT})$ and
the coefficient $A$ is different. Indeed, using the same notation as
before, one obtains
\bea
\label{al}
\frac{\tilde{m}_{\tau}(M_{GUT})}{A} = \frac{\sqrt{ a + \xi_N^2 b}}{\xi_N}
\eea
Analogously to the charged-lepton sector, the down-quark sector at low 
energies is now given by
\bea
m_{D}(\mu) \simeq 
 \left (
 \begin{array}{cc}
0  & 0 \\
I_t^{1/3} \xi_t x & I_t^{1/3} \xi_t 
 \end{array}
 \right).
\eea
and therefore the left-handed mixing angle is equal to zero, while
the extrapolated value of the bottom Yukawa coupling at $M_{GUT}$,
$\tilde{m}_b(M_{GUT})$, is related to the (33) element of the 
down quark mass matrix by
\bea
(m_D)_{33}(M_{GUT}) = \tilde{m}_b(M_{GUT})
\frac{(m_D)_{33}}{m_b} \equiv A. 
\eea
or, equivalently,
\bea
\label{ad}
(m_D)_{33}(M_{GUT}) = \frac{\tilde{m}_b}{\sqrt{x^2 + 1}} \equiv A. 
\eea
Equating the values of $A$ in Eqs. (\ref{ad}) and (\ref{al}), we 
obtain
\bea
\frac{\tilde{m}_{\tau}(M_{GUT})}{\tilde{m}_b(M_{GUT})} = 
\frac{1}{\xi_N} \sqrt{ \frac{a + \xi_N^2 b}{1 + x^2} }
\eea
Finally, for the values of $x$ and $y$ which fulfill the condition 
(\ref{ymaxmix}) of maximal mixing, we obtain
\bea
\label{comparison1}
\frac{\tilde{m}_{\tau}(M_{GUT})}{\tilde{m}_b(M_{GUT})} =
\sqrt{\frac{2}{1 + \xi_N^2}}.
\eea
The main difference between Eqs. (\ref{comparison}) and (\ref{comparison1})
resides in the fact that the factor $(1 + x^2)$ does not appear in
the latter. This is just a reflection of the structure of the
down and lepton quark masses at the $GUT$ scale. Moreover, 
(\ref{comparison1}) is independent of $x$, although it implicitly
assumes that the condition (\ref{ymaxmix}) for maximal mixing is
fulfilled. Since the factor $\sqrt{2/(1+\xi_N^2)} \leq 1/\xi_N$, 
it follows from 
Fig.~1 that approximate unification can be achieved only for 
values of $\tan\beta$ consistent with the infrared fixed-point solution
for
the top-quark mass. Moreover, for the condition of unification of the
top-quark and neutrino Yukawa couplings
at the top quark mass fixed point solution, for which $\xi_N \simeq
0.87$, unification can only be achieved for  
somewhat larger values of the bottom 
mass (or lower values of $\alpha_3(M_Z)$) than the one used in Fig.~1,
namely $M_b \simeq 5.1$ GeV corresponding to values of 
$m_b(M_b) \simeq 4.5$--4.6 GeV.
These values of the bottom mass, although high, are still consistent
with phenomenological constraints. In this sense, the situation is
better than the case where there is
no mixing in the lepton sector, for which unacceptable values of 
$m_b(M_b) \simeq 4.8$--4.9 GeV would be required in order
to achieve $b-\tau$ and $t-\nu_{\tau}$ 
unification~\cite{VS,Rat}. 

\section{Conclusions}

In this paper, we analyzed quark-lepton mass unification
in the light of the evidence of neutrino
oscillations coming from the recent Super-Kamiokande data. 
We mainly concentrated  on $b-\tau$ Yukawa coupling unification,
but have also examined $t-\nu$ Yukawa unification in the
small and moderate $\tan\beta$ regime of the MSSM.
We have
shown that, in the case that there is only small right-and left-handed
mixing in the down-quark sector, Yukawa 
coupling unification can be achieved
for values of $\tan\beta$ larger than those consistent with the
infrared fixed-point solution of the top-quark mass. Although,
for simplicity, we analysed the case of symmetric mass
matrices at the scale $M_{GUT}$, this result can be easily
generalized to the general case of non-symmetric lepton mass matrices,
leading to a large hierarchy between the Dirac mass eigenvalues. On the 
other hand, if exact $SU(5)$ relations between the right-handed down-quark
and left-handed charged-lepton mass matrices are used, Yukawa
coupling unification can be achieved only for small values of
$\tan\beta \simeq 1.5$, close to the infrared fixed-point solution for
the top-quark mass.

These results cast in a new light the theoretical interest of
intermediate values of $\tan\beta$, which had previously been 
disfavoured on the basis of Yukawa coupling unification arguments. We 
now see that there is a subtle interplay between these arguments and
neutrino masses and mixing, which opens up new possibilities at
intermediate $\tan\beta$. 

One significant implication of these results
is for Higgs searches
at LEP. In the absence of lepton mixing, 
for small values of $\tan\beta$ 
consistent with $b-\tau$ mass unification, values of the Higgs mass in
the unexcluded range 95--105 GeV still accessible to LEP could be obtained
only for
large values of the stop masses
and of the stop mixing parameter~\cite{CCPW}.
On the other
hand, at large $\tan\beta$, the Higgs mass tends naturally to values which
are beyond this range, unless one of the stops is relatively light
and the stop mixing parameter is small (as happens, for instance,
in scenarios consistent with electroweak baryogenesis~\cite{QCW}). 
The range  $95 \leq m_h \leq 105$~GeV of Higgs masses is 
achieved most naturally
for intermediate values $2 \leq \tan\beta \leq 4$.
Therefore, finding the a Higgs boson in this mass range at LEP would
probably hint towards either a light stop, or such
intermediate values of $\tan\beta$. As we have shown
in this article, $b-\tau$ Yukawa coupling
unification can still be achieved for $2 \leq \tan\beta \leq 4$
if the mass-matrix textures at $M_{GUT}$
are such that the left- and right-handed down-quark  mixing is small, 
whereas the expected large mixing in the lepton sector is
shared between the neutrino and the charged-lepton sector, with
a significant contribution from the latter. Thus, there is an
intriguing theoretical linkage between 
neutrino physics and the Higgs search at LEP, which passes via
mass unification.

\vskip 1. cm
~\\
{\bf Acknowledgements} We thank G.F. Giudice for useful discussions. 
C.W. would also like to thank S. Raby for very stimulating discussions.
Work supported in part by the US Department of Energy, Division of High
Energy Physics, under Contract W-31-109-ENG-38.

\newpage
{\large \bf Appendix A. Lowering $\tan\beta$ in $SO(10)$ Models}

In this appendix we present a mechanism to lower the predicted
value of $\tan\beta$ in the framework of SO(10) models,
like the ones considered in this article.
Low values of $\tan\beta$ can easily be achieved by assuming that
only one ${\bf 10}$ of GUT Higgs fields couples to fermions, but 
that this ${\bf 10}$
contains only some components of the two electroweak Higgs doublets, the
other
components coming, for instance, from an additional ${\bf 10}$~\cite{CMS}.
The overall effect is
to multiply the down and lepton mass matrices by a factor
$\omega$, which is the ratio of the relative components of the
two Higgs doublets in the ${\bf 10}$ which couples to fermions. The 
minimal model would hence be obtained for $\omega = 1$. Here,
in order to obtain $\omega > 1$, we may
implement a slightly modified version  of the mechanism implemented
in~\cite{CMS}, as we now describe.

Consider the superpotential
\begin{equation}
W = {\bf 10} \; {\bf 45}_{B-L} {\bf 10}^{'}  + 
\left[ M_1 \; {\bf 10} + \left( M_2 + {\bf 45}_X \right) 
{\bf 10}^{'} \right] {\bf 10}^{''},
\end{equation}
where $M_1$ and $M_2$ are of order $M_{GUT}$, ${\bf 10}$,
${\bf 10^{'}}$ and
${\bf 10^{''}}$ are decuplets, and only ${\bf 10}$ participates in the
fermion mass operators.

The first term in $W$ implements the $SO(10)$ missing-doublet
mechanism~\cite{DW}, and yields 4 light doublets:
{ 2}, ${\bf \bar{2}}$, ${\bf 2^{'}}$ and ${\bf \bar{2}^{'}}$,
whereas the corresponding color triplets 
acquire a mass of order $M_{GUT}$.
The second term gives a mass to a linear combination of
{ 2} and ${\bf 2^{'}}$, by pairing it with ${\bf \bar{2}^{''}}$,
and a different linear combination of ${\bf \bar{2}}$ and
${\bf \bar{2}^{'}}$, by pairing it with ${\bf 2^{''}}$.
Explicitly, the light states are given by
\begin{equation}
{\bf 2}_L = \frac{ M_1 \; {\bf 2^{'}} - \left( M_2 + {\bf v} \right)
{\bf 2}}{\sqrt{M_1^2
+ \left(M_2 + {\bf v} \right)^2}}
\end{equation}
and
\begin{equation}
{\bf \bar{2}}_L = \frac{ M_1 \; {\bf \bar{2^{'}}}  - 
\left( M_2 - {\bf v} \right) {\bf \bar{2}}}
{\sqrt{M_1^2 + \left(M_2 - {\bf v} \right)^2}} ,
\end{equation}
where $<{\bf 45}_X> = {\bf v} \times X$, with $X = +$ $(-)$ when it acts
on the $SU(5)$ ${\bf 5}$  (${\bf \bar{5}}$) components
of a ${\bf 10}$ representation,
respectively.  Since ${\bf 2}$ couples to the up quarks and
${\bf \bar{2}}$ couples to the down quarks, in this example we have
\begin{eqnarray}
\lambda_t & = & \lambda \frac{M_2 + {\bf v}}{\sqrt{M_1^2 + 
\left(M_2 + {\bf v} \right)^2}} ,
\nonumber\\
\lambda_b & = & \lambda \frac{M_2 - {\bf v}}{\sqrt{M_1^2 + 
\left(M_2 - {\bf v} \right)^2}} ,
\end{eqnarray}
and
\begin{equation}
\omega = \frac{M_2 - {\bf v}}{M_2 + {\bf v}} \times
\frac{\sqrt{M_1^2 + \left(M_2 + {\bf v} \right)^2}} 
{\sqrt{M_1^2 + \left(M_2 - {\bf v} \right)^2}}.
\end{equation}
Notice that, in this simple example, 
$M_1$  cannot be too small, 
or else  a pair of light triplets  
would appear in the spectrum, affecting the prediction for
$\sin^2 \theta_W$. This restriction does not apply
to $M_2 - {\bf v}$, as long as $M_2 + {\bf v}$ is still of the order
of the GUT scale. \\

\vskip 0.5 cm

~\\
{\large \bf Appendix B.  On the Mixing in the Heavy Majorana Sector}

In this appendix we generalize the analysis presented in this
article, for the general case of hierarchical Dirac mass
matrices for charged leptons and neutrinos, and non-vanishing
mixing in the heavy Majorana neutrino sector.
Considering a symmetric mass matrix for the
Majorana neutrinos at the GUT scale, of the form
\bea
{\cal M}_{N}(M_{GUT}) = M
 \left (
 \begin{array}{cc}
f & g \\
g & h 
 \end{array}
 \right),
\eea
we transform it to the basis in which the Dirac neutrino-mass matrix
is diagonal. In this basis, the Majorana mass of the neutrinos will
take a form
\bea
{\cal M}_{N}'(M_{GUT}) = M
 \left (
 \begin{array}{cc}
f' & g' \\
g' & h' 
 \end{array}
 \right).
\label{thism}
\eea
where we do not write out the formal expressions of the
matrix  elements as
functions of $f,g$ and $h$, since they are not essential for our
discussion. In the same basis, the Dirac mass matrix of the neutrinos
is given approximately by $m_{\nu}^D = {\rm diag}( 0, B ( 1 + y^2))$.
The evolution of the mass matrix, Eq. (\ref{thism}) to the relevant 
Majorana mass scale leads to 
\bea
{\cal M}_{N}'(M_N) = M
 \left (
` \begin{array}{cc}
f' & g'\xi_N^2 \\
g'\xi_N^2 & h' \xi_N^4 
 \end{array}
 \right).
\eea
where the scale $M_N$ should be identified with  
the inverse of $({\cal M}^{-1}_N)_{33}$.
If we now compute the effective low-energy Majorana mass for the
left-handed neutrinos, we obtain
\bea
m_{eff}'(M_{N}) = \lambda_{N}^2 H_2^2 \cdot ({\cal M}_N^{-1})_{33}
 \left (
 \begin{array}{cc}
0 & 0 \\
0 & 1 
 \end{array}
 \right).
\eea
with 
\bea
({\cal M}_N^{-1})_{33} = \frac{f'}{(f'g'-h'^2)\xi_N^4}.
\eea
The important thing to notice is that, as in the case of 
a unit Majorana neutrino-mass matrix, no mixing in the
neutrino sector is induced.
So, in the case of hierarchical
Dirac neutrinos, the mixing in the heavy Majorana sector
does not affect the observed lepton mixing at
low energies \cite{MG}.

We observe that the above conclusion was obtained by approximating
the smallest neutrino mass to zero. In the more realistic case
of a small but non-vanishing neutrino mass, the above conclusion
holds unless $h' \gg f',g'$, since then the approximations
made in the above analysis would be invalid. Therefore,
the analysis presented in this article
holds in the quite general case considered in
this article, in which the hierarchy of neutrino masses is 
induced by the Dirac mass structure and the elements of the 
heavy Majorana mass matrix are of the same order 
(with ${\rm det}{\cal M}_N \neq 0$),
independently of the mixing in the Majorana mass sector.

\end{document}